\documentclass[conference,10pt]{IEEEtran}
\IEEEoverridecommandlockouts
\pdfoutput=1
\usepackage{cite}
\usepackage{amsmath,amssymb,amsfonts}
\usepackage{algorithmic}
\usepackage{graphicx}
\usepackage[tight]{subfigure}
\usepackage{acronym}
\usepackage{textcomp}
\usepackage{siunitx}
\usepackage{xcolor}
\usepackage{bm}
\def\BibTeX{{\rm B\kern-.05em{\sc i\kern-.025em b}\kern-.08em
    T\kern-.1667em\lower.7ex\hbox{E}\kern-.125emX}}

\begin{document}

\title{Approximate Message Passing for Indoor THz Channel Estimation}

\author{Viktoria Schram, Ali Bereyhi, Jan-Nico Zaech, Ralf R. M\"uller  and Wolfgang H. Gerstacker\\
Institute for Digital Communications, Friedrich-Alexander University Erlangen-N\"urnberg (FAU)\\
Cauerstr. 7, D-91058 Erlangen, Germany \\ \{viktoria.schram, ali.bereyhi, jan-nico.zaech, ralf.r.mueller, wolfgang.gerstacker\}@fau.de}

\maketitle

\begin{abstract}
Compressed sensing (CS) deals with the problem of reconstructing a sparse vector from an under-determined set of observations. Approximate message passing (AMP) is a technique used in CS based on iterative thresholding and inspired by belief propagation in graphical models. Due to the high transmission rate and a high molecular absorption, spreading loss and reflection loss, the discrete-time channel impulse response (CIR) of a typical indoor THz channel is very long and exhibits an approximately sparse characteristic. In this paper, we develop AMP based channel estimation algorithms for indoor THz communications. The performance of these algorithms is compared to the state of the art. We apply AMP with soft- and hard-thresholding. Unlike the common applications in which AMP with hard-thresholding diverges, the properties of the THz channel favor this approach. It is shown that THz channel estimation via hard-thresholding AMP outperforms all previously proposed methods and approaches the oracle based performance closely.
\end{abstract}


\section{Introduction}
Terahertz (THz) communication is considered as a promising paradigm to support ultra-high data rates and large bandwidths to meet the ever-increasing demand for high capacity wireless communications. This emerging technology  has the potential to transform the communications landscape and  revolutionize the way how  people create, share and consume information \cite{elayan2018terahertz}. For its realization, various technology options can be thought of as THz frequencies are bridging the gap between the mmWave and infra-red frequencies and therefore foster hardware developments that are either based on electronics or optics \cite{elayan2018terahertz}. Various ongoing research and innovation programs on both fields, for instance, Horizon2020 funded projects like Terranova, EPIC or Terapod, and continuous improvements indicate that THz communications is highly likely to become operational for future beyond 5G systems \cite{horizon2020}.  
\par
Earlier studies have demonstrated that data rates in the order of several hundred \si{Gbps} or even \si{Tbps} can indeed be achieved \cite{moldovan2016globecom,moldovan2016data,schramFB,akyildiz2014terahertz,jornet2011channel,akyildiz2014teranets}. The results in~\cite{moldovan2016globecom} verify such rates for indoor transmission distances of up to $ 1\,\si{m} $. In \cite{moldovan2016data} different bounds for the maximum tolerable distance for a prescribed performance are proposed, considering various transmission schemes and an efficient power allocation algorithm. The investigation of finite blocklength coding for THz communications in \cite{schramFB} confirms the achievability of transmission rates up to $110$ \si{Gbps}. These promising theoretical results along with steady improvements in  hardware realizations encourage further development of \si{THz} communication systems. 
\par 
The \si{\THz} channel  suffers from  a high molecular absorption which leads to signal distortion and colored noise, cf. e.g.~\cite{moldovan2016globecom}.  Additionally, there occur losses due to spreading, which result in a very high and frequency-selective path loss. Such a behavior affects even line-of-sight (LOS) links of \si{\THz} transmission~\cite{moldovan2016globecom}. The non-line-of-sight (NLOS) components are further extremely influenced by high reflection losses depending on the characteristics of the surfaces reflecting the \si{\THz} waves~\cite{moldovan2016globecom}. Such difficult transmission characteristics limit \si{\THz} wave propagation to a few meters~\cite{moldovan2016globecom}  and lead to a sparse equivalent discrete-time channel impulse response (CIR); see \cite{schramCS} for more details.
\par 
In order to fully exploit  the sparsity of the THz channel, CS techniques can be employed for channel estimation \cite{schramCS}.  CS deals with the problem of  recovering samples of sparse signals from an underdetermined set of measurements. For sake of computational tractability,  classical CS approaches are mainly based on convex optimization techniques \cite{tibshirani1996regression, chen2001atomic}.
Despite the tractable complexity of classical CS approaches, the direct implementation of these techniques via linear programming is still considered complex in applications with large dimensions \cite{donoho2009message}. This issue is resolved by developing iterative algorithms whose complexity scales linearly with the signal space dimension \cite{foucart,MalekiCS}. Conventional iterative approaches often exhibit linear complexity at the expense of a considerable performance degradation. Such a phenomenon can be also observed in CSbased THz channel estimation; see our earlier study in \cite{schramCS}. Nevertheless, for signals with large dimensionality, this degradation can be avoided by using  approximate message passing (AMP) algorithms. AMP algorithms recover sparse signals by approximating the update rules of the sum-product algorithm for the original convex optimization problem in the large-system limit. These approximations, which yield linear computational complexity, are shown to be asymptotically accurate. Hence, in the asymptotic regime, classical CS performance is achieved via AMP with linear complexity. \cite{donoho2009message}. 
\par 
In this work we investigate AMP recovery approaches for THz channel estimation and compare them to conventional CSbased iterative approaches.
\section{Related Work}
The original AMP algorithm has been extensively refined since its first introduction in \cite{donoho2009message}. Two popular extensions are generalized approximate message passing (GAMP) \cite{GAMP} which is applicable to a large class of non-Gaussian estimation problems and vector approximate message passing (VAMP) \cite{VAMP} which extends the requirement of having i.i.d. sub-Gaussian sensing matrices to unitarily invariant sensing matrices.\par
These concepts have  been further developed and their advantages have been leveraged for a wide range of channel estimation tasks in different areas. For instance,  a parametric bilinear GAMP algorithm is proposed for joint channel estimation and equalization in \cite{sun2018joint}, where AMP is extended to a quantized bilinear model, considering single-input single-output equalization of single-carrier block transmission over fading channels. The authors in \cite{zhang2018one} generalized this result to one-bit quantized massive MIMO receivers and proposed a variational AMP which uses variational approximation to deal with non-linearities. The proposed scheme is able to accomplish joint channel estimation and data detection and is shown to outperform both linear detectors and detectors based on variational Bayesian inference. In \cite{borgerding2017amp} the authors propose learned AMP and VAMP networks combined with deep learning offering a better robustness against deviation of the statistics of the sensing matrix from an i.i.d. Gaussian distribution.
\par 
The remainder of this paper is organized as follows. In Section \ref{III}, the system model and the relevant fundamentals of AMP are introduced. In Section \ref{IV}, an evaluation of the considered schemes is presented, including a qualitative analysis of AMP in Section~\ref{sub1} and a discussion of numerical results in Section \ref{sub2}.  Finally, Section~\ref{IV} concludes the paper.
\section{System Model and Fundamentals}\label{III}
In this section the THz transmission model is introduced and a short introduction to relevant fundamentals of AMP is presented.

\subsection{System model}
We consider the \emph{indoor} \si{\THz} communication channel model, obtained by using ray tracing and noise model as given in \cite{schramCS}. This model results in a highly frequency-selective  channel transfer function $ H_{\mathrm{eq}}\left(f,r,\bm{\zeta} \right) $, where $f$ refers to the frequency,  $r$ denotes the distance between the transmitter and the receiver, and vector $\bm{\zeta}$ is composed of environmental parameters. As in \cite{schramCS}, we assume a subband transmission approach, i.e., the total bandwidth $ BW $ from $0.1\,\si{THz}$ to $1\,\si{THz}$  is divided into $ N $ subbands of equal width $ \Delta{f} =\frac{BW}{N}$ such that the dispersion of each resulting subchannel is better controlled. In the sequel we assume that a particular subband has been selected for the transmission. Each subchannel is modeled to be time-invariant during a certain transmission interval (transmit signal burst).  It should be noted that a subband approach is better realizable since transmission over the full bandwidth faces significant challenges regarding device technology~\cite{moldovan2016data}.
\par
In channel estimation, the estimate $\hat{\mathbf{h}} \in \mathbb{C}^n$ of the channel vector of length $n$ can be obtained from the received signal $\mathbf{y} \in \mathbb{C}^m$ which reads
\begin{equation}\label{toeplitz}
\mathbf{y}=\mathbf{A} \mathbf{h}+\mathbf{n},
\end{equation}
with $\mathbf{n}\in\mathbb{C}^{m}$ denoting the noise vector, $m$ being the training sequence length and $\mathbf{A}\in\mathbb{R}^{m\times n}$ representing a Toeplitz-structured convolution matrix with binary phase-shift keying (BPSK) symbols; see~\cite{schramCS} for more details. $\mathbf{A}$ is also called the \emph{sensing matrix}, as it is used to probe the channel. The channel vector $\mathbf{h}$ comprises all $n$ channel taps, i.e., $h[l]$, $l=1,...,n$.

\subsection{Basics of Sparse Recovery}
To meet the requirements of standard AMP algorithms, we modify the sensing matrix $\mathbf{A}$ such that it fulfills the constraint of unitary invariance. To this end we normalize the matrix such that each column has unit $l_2$-norm \cite{bajwa2009new}. \par 
For a sparse recovery problem, the indeterminacy degree is defined as \cite{maleki2009optimally}
\begin{equation}
\delta=\frac{m}{n},
\end{equation}
and the corresponding compression rate is given by $r=\delta^{-1}$. High compression rates are of particular interest as they occur when probing the THz channel exhibiting a long impulse response with a limited number of training symbols, corresponding to $\delta<<1$. Without any prior information on signal statistics, $m\geq n $ measurements are required to reconstruct $\mathbf{h}$ by using the conventional least squares approach. In CS, however, $\mathbf{h}$ is known to be sparse, i.e., only $k\leq n$ entries of $\mathbf{h}$ are non-zero, i.e., $||\mathbf{h}||_0 \leq k$. The sparsity factor for such a signal is defined as $\rho=\frac{k}{n}$ \cite{maleki2009optimally}.
Considering the theoretical lower bound  $m \geq k$, we introduce the normalized sparsity as \cite{maleki2009optimally}
\begin{equation}
\rho'=\frac{k}{m}.
\end{equation}
\subsection{Fundamentals of AMP}
\begin{figure}[t!]
	\centering
	\subfigure[][]{
	\label{subfig:softTh}
   	    \includegraphics[width=0.19\textwidth, trim= 25mm 0mm -7mm 20mm]{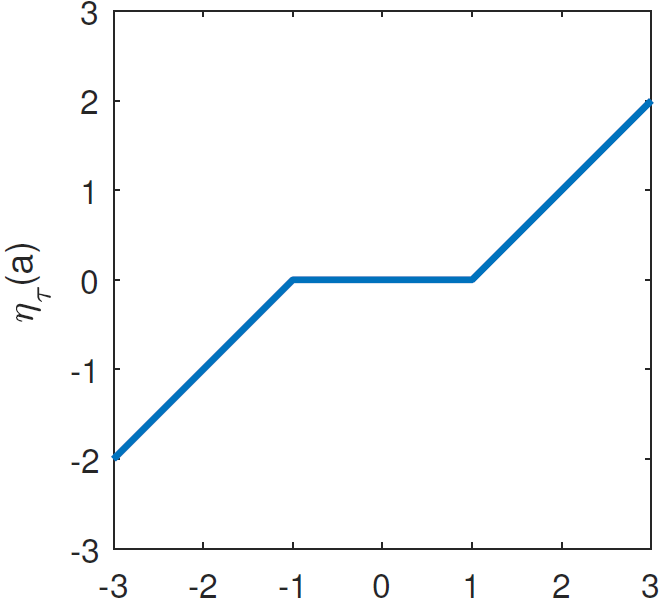}}
   	    \hspace{5pt}
	\subfigure[][]{
		\label{subfig:hardTh}
   	    \includegraphics[width=0.183\textwidth, trim= 25mm 0mm 0mm 4mm]{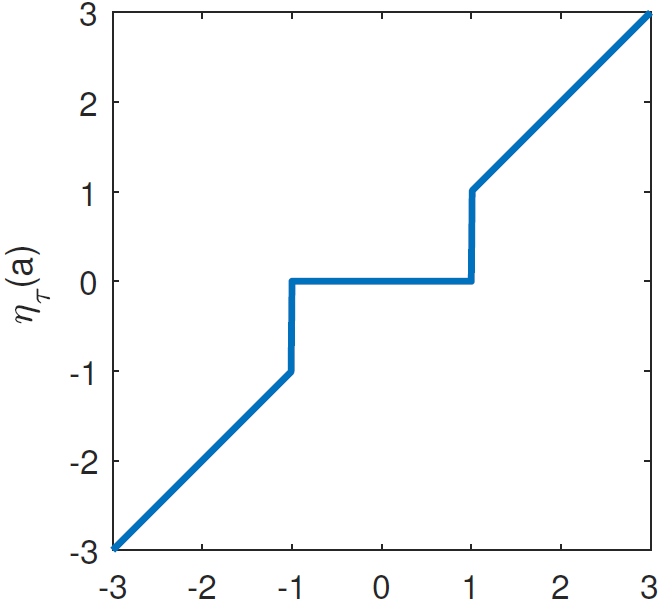}}
   	    \caption[]{
   	    \subref{subfig:softTh}~Soft-thresholding function.
   	    \subref{subfig:hardTh}~Hard-thresholding function.\\ $\eta_{\tau}(a)$ is plotted over $a$.}
   	    \label{fig:denoiser}
   	  
	\end{figure}

A generic AMP algorithm consists of a 
\textit{nonlinear} and a \textit{linear} step which are as follows \cite{GAMP}
\begin{align}
{\hat{\mathbf{h}}}^{t+1}&=\eta_\tau (\mathbf{A}^\text{H}{\mathbf{z}}^t+{\hat{\mathbf{h}}}^t), \\
{\mathbf{z}}^t &= \mathbf{y} - \mathbf{A}{\hat{\mathbf{h}}}^t+\frac{1}{\delta}{\mathbf{z}}^{t-1}\langle \eta'_\tau(\mathbf{A}^\text{H}\mathbf{z}^{t-1}+\hat{\mathbf{h}}^{t-1})\rangle,
\end{align}
where $\mathbf{A}^\text{H}$ denotes the Hermitian transpose of $\mathbf{A}$ and $\langle \cdot \rangle$ is the expectation operator. The expression $\langle \eta'_\tau(\mathbf{A}^\text{H}\mathbf{z}^{t-1}+\hat{\mathbf{h}}^{t-1})\rangle$ is referred to as the \textit{Onsager correction term} with $\eta'_\tau ( \cdot )$ being the derivative of $\eta_\tau$, which is a nonlinear function that imposes the sparse structure of the signal on the estimate.  The integer $t$ denotes the iteration index, and $\hat{\mathbf{h}}^t$ and $\hat{\mathbf{z}}^t$ are the estimate and the current residual in iteration $t$, respectively. There are several possible choices for $\eta_{\tau} ( \cdot )$ \cite{metzler2016denoising}, i.e., soft-thresholding given by \cite{donoho2009message}
\begin{equation}\label{6}
\eta^{\text{soft}}_\tau(a)=\begin{cases}
a-\tau\sigma & \text{if  }{\;} a{\;}\geq \tau\sigma\\
0 & \text{if } |a|\leq \tau\sigma \\
a+\tau\sigma &\text{if  } {\;}a{\;} \leq -\tau\sigma 
\end{cases},
\end{equation}
and hard-thresholding realized by \cite{blumensath2007iterative}
\begin{equation}\label{7}
\eta^{\text{hard}}_\tau(a)=\begin{cases}
a & \text{if } |a|\geq \tau\sigma\\
0 & \text{if } |a|\leq \tau\sigma 
\end{cases}.
\end{equation}
in \eqref{6} and \eqref{7}  $\sigma$ is the standard deviation of the residual in the current iteration. The value of $\sigma^2$ in iteration $t$ is given by \cite{metzler2016denoising}
\begin{equation}
\sigma_t^2\approx||\mathbf{z}^t||_2^2/m.
\end{equation}
Both hard- and soft-thresholding functions are shown in Fig.~\ref{fig:denoiser}, where $\tau$ and $\sigma$ are set to one.\par 
\subsection{Regularized Least Squares Schemes Corresponding to Soft- and Hard-Thresholding}
The regularized least squares (RLS) scheme with $\ell_0$-norm regularization  for channel estimation reads \cite{foucart2017mathematical}
\begin{equation}\label{l0}
\hat{\mathbf{h}}=\underset{\tilde{\mathbf{h}}} {\mathrm{argmin}}\frac{1}{2\lambda}  ||\mathbf{y}-\mathbf{A}\tilde{\mathbf{h}}||_2^2+||\tilde{\mathbf{h}}||_0
\end{equation}
Here, $\lambda$ is the so-called regularizer and needs to be adjusted w.r.t. the signal sparsity factor $\rho$. This problem is an NP hard optimization problem. By using convex relaxation, the recovery scheme in (9) reduces to the least absolute shrinkage and selection operator (LASSO) scheme which reads~\cite{tibshirani1996regression}
\begin{equation}\label{l2}
\hat{\mathbf{h}}=\underset{\tilde{\mathbf{h}}} {\mathrm{argmin}}\frac{1}{2} ||\mathbf{y}-\mathbf{A}\tilde{\mathbf{h}}||_2^2+\lambda||\tilde{\mathbf{h}}||_1
\end{equation}
and is also known as basis pursuit denoising (BPDN). 
AMP with hard-thresholding (H-AMP) results in an approximate solution for \eqref{l0} while AMP with soft-thresholding (S-AMP) yields a good approximation for the solution of \eqref{l2} \cite{starck2009overview}.
\par
When reconstructing a signal from its measurements with indeterminacy $\delta$, any recovery approach will fail if the sparsity factor $\rho$ is increased beyond an algorithm-specific value. For a finite system size, the number of successful recoveries will decrease smoothly with growing sparsity factor, but the transition will get steeper for increasing system size. Eventually, in the large-system limit, the boundary between successful and failed recoveries gets sharp and is called phase transition \cite{MalekiCS}. The phase transition can be visualized by depicting the probability of successful reconstruction over the plane spanned by the indeterminacy $\delta$ and the normalized sparsity $\rho'$ \cite{MalekiCS}. An optimum reconstruction approach should be able to reconstruct the signal vector as long as $m \geq k$, which corresponds to the complete area corresponding to $0 \leq \rho ' \leq 1$, $0 \leq \delta \leq 1$. \par
\begin{figure}[t!]
    	\center{
   	    \includegraphics[width=0.3\textwidth, trim= 20mm 0mm 0mm 0mm]{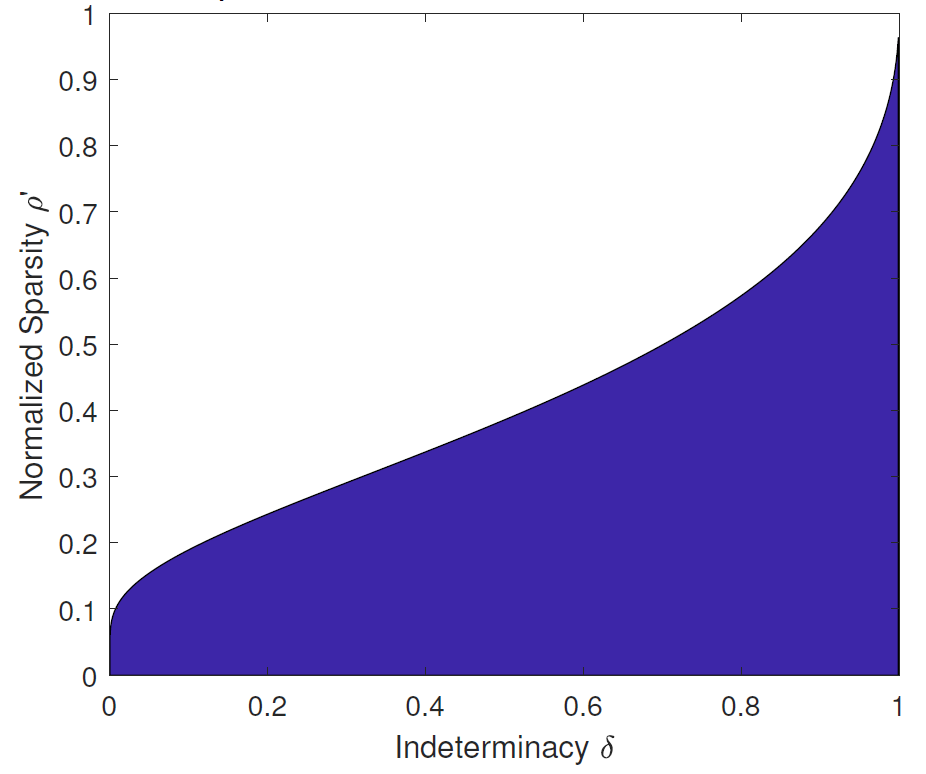}}
   	    \caption{Phase transition diagram for $l_1$-norm minimization. Blue: successful reconstruction.
   	    \label{fig:phaseTrans}}
   	    \vskip -5mm
	\end{figure}

 
The phase transition for the $l_1$-norm minimization is depicted in Fig.~\ref{fig:phaseTrans}, with tabulated values according to \cite{tann}. The sharp phase transition holds in the large-system limit, i.e., $n \rightarrow \infty$. For low compression rates $(r  \rightarrow 1)$, the convex relaxation performs well, while it degrades severely for high compression rates $(r \rightarrow \infty)$. This behavior agrees with the intuition that the $l_1$-norm is a coarse approximation of the $l_0$-norm and does not measure the exact sparsity \cite{tann}. 

\section{Performance Evaluation}\label{IV}
In this section, we first discuss the convergence of the H-AMP and S-AMP algorithms.  The performance of these algorithms for THz channel estimation is then evaluated and compared to that of the methods analyzed in \cite{schramCS}. {For reconstruction, sensing matrices that satisfy necessary conditions for being used in an AMP framework are assumed, namely being either Gaussian matrices with independent and identically distributed (i.i.d.) entries or subsampled unitarily invariant matrices \cite{bajwa2009new}. The sensing matrix in the latter case is constructed by normalizing $\mathbf{A}$ such that each column has unit $l_2$-norm \cite{bajwa2009new}.}
\subsection{Convergence properties of H-AMP and S-AMP}\label{sub1}
For AMP, only the threshold $\tau$ is a free parameter that needs to be optimized according to the signal sparsity. The optimal value of $\tau$ minimizes the normalized mean squared error ($\text{NMSE}$) of the reconstructed signal vector, defined as
\begin{equation}
\text{NMSE}=\frac{||\hat{\mathbf{h}}-\mathbf{h}||_2^2}{||\mathbf{h}||_2^2},
\end{equation}
after a sufficient number of iterations.  \par
\begin{figure}[t!]
    	\center{
   	    \includegraphics[width=0.5\textwidth, trim= 20mm 0mm 0mm 0mm]{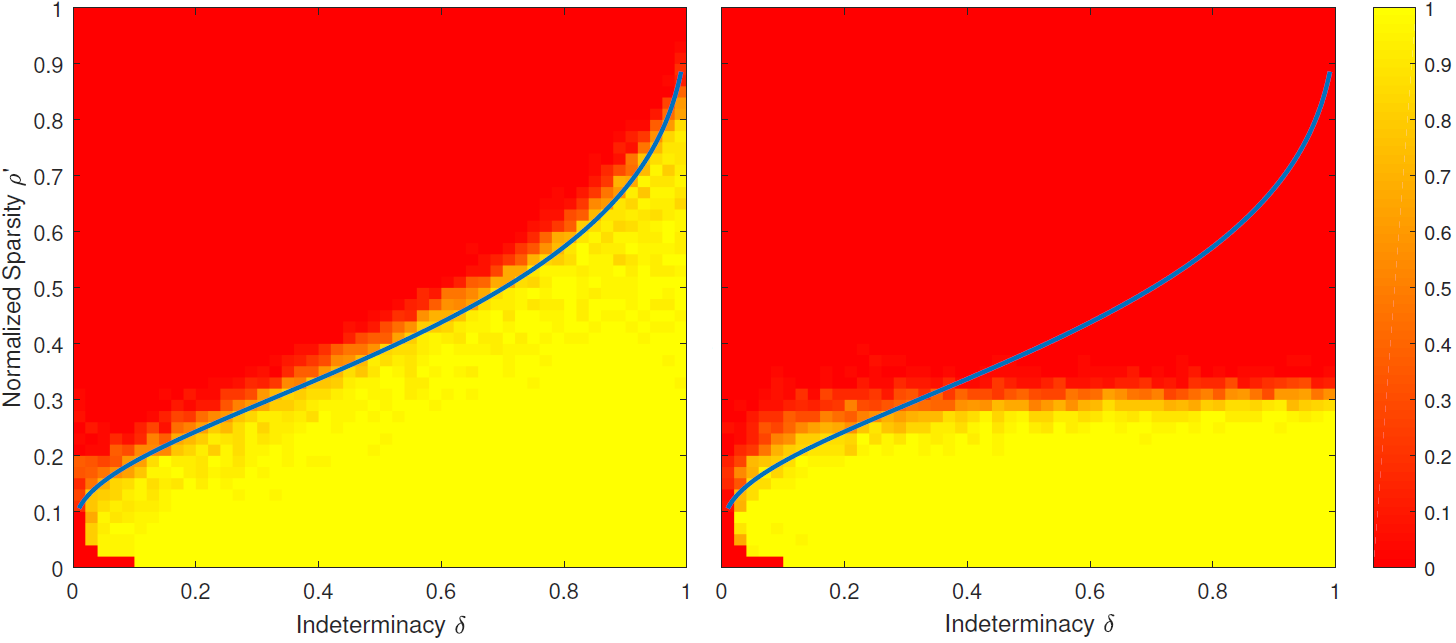}}
   	    \caption{Empirical phase transition diagram for S-AMP (left), H-AMP (right) and  $l_1$-norm minimization (blue line). Yellow: successful reconstruction.\label{fig:trans}}
	\end{figure}

{To measure the probability of proper reconstruction, multiple randomly drawn compressed sensing problems were considered for each parameter set ($n,m,\rho,\sigma$), i.e, randomly generated strictly sparse $\mathbf{h}$ were estimated by using i.i.d. Gaussian matrices as sensing matrices. One of the most important performance characteristics is the phase transition.} The threshold to accept a reconstruction as successful is set to $\text{NMSE}=-20$ dB, as this corresponds to estimation with high fidelity and leads to a steep phase transition. For the optimal selection of $\tau$ we use oracle and further optimization is possible. To measure the performance of H-AMP and compare it to that of S-AMP, we consider various scenarios and study the respective phase transition.  
Our first analysis was performed for a signal vector of length $n=1000$. Here, the sparsity $\rho$ and the compression rate $r$ were varied from $0.01$ to $0.99$ with a step size of  $0.02$. AMP is terminated after $200$ iterations and $30$ measurements have been taken for every parameter set.  An overview of the empirical phase transitions of S-AMP and H-AMP along with the analytical phase transition of $l_1$-norm regularization is shown in Fig.~\ref{fig:trans}. The colored patches indicate the probability of successful reconstruction for a specific configuration of $\delta$ and $\rho'$, and the blue line is the analytical phase transition corresponding to a $l_1$-norm minimization. 
	
	\begin{figure}[t!]
    	\center{
   	    \includegraphics[width=0.5\textwidth, trim= 20mm 0mm 0mm 0mm]{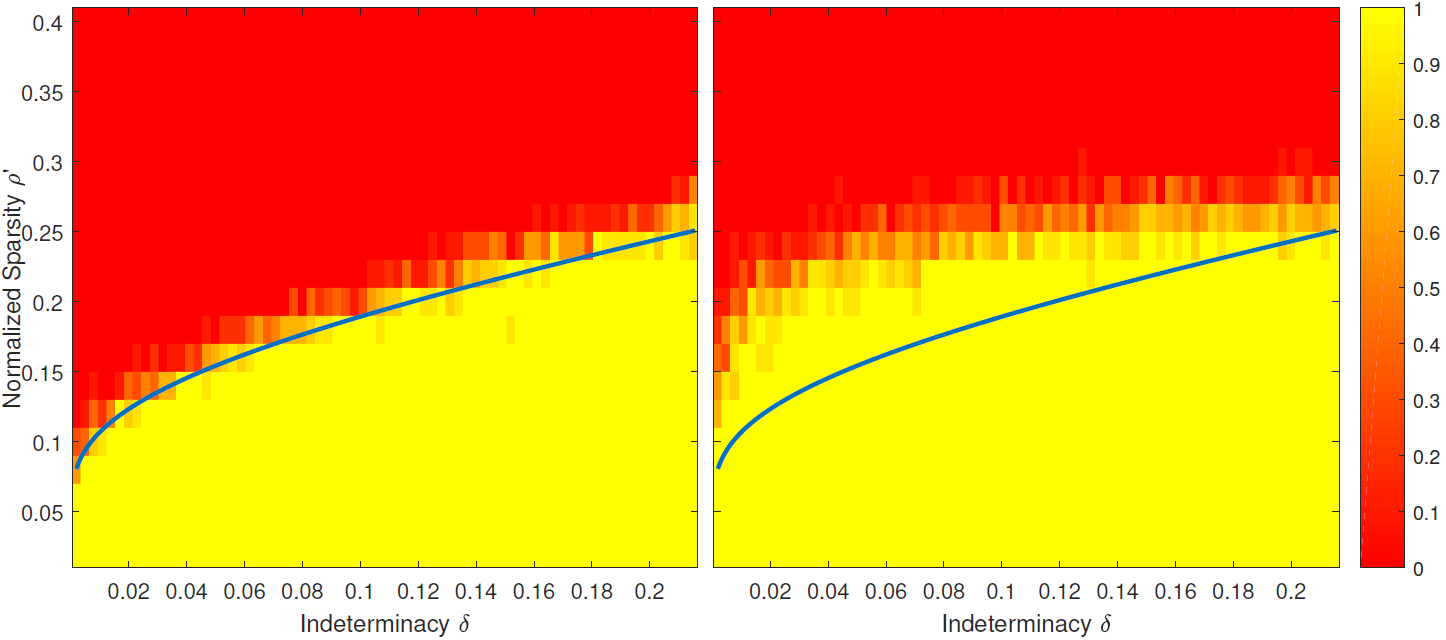}}
   	    \caption{Empirical phase transition diagram for S-AMP (left), H-AMP (right)  and  $l_1$-norm minimization (blue line) for low indeterminacy. Yellow: successful reconstruction.\label{fig:translow}}
	\end{figure} 
Comparing the empirical phase transition for S-AMP and $l_1$-norm minimization shows a very good agreement for the region of success. This was expected as it was shown analytically in \cite{donoho2009message} that both methods are characterized by the same phase transition. However, for H-AMP the reconstruction performance lies below that of S-AMP. For indeterminacy values of
$\delta $ larger than $ 0.2$ the algorithm often converges to incorrect solutions. One can observe that a frequent convergence to an
incorrect solution is related to a high probability of an all-zero reconstruction vector. Furthermore,
the phase transition gets smeared for such low indeterminacy values, leading to the conclusion
that the problem size obtained by the parameter settings is too low in those cases. Setting the parameters to e.g. $\delta = 0.2$ and
$\rho = 0.1$ leads to $m\rho = 20$ nonzero elements for the given measurement size of $n = 1000$. This
small number of measurements does not correspond to a large system, which renders the assumptions
used during derivation of AMP incorrect and therefore does not allow statements about the
algorithm's properties for low indeterminacy values.\par
	\begin{figure}[t!]
	\vspace{6mm}
	\centering
	\subfigure[][]{
	\label{subfig:1}
   	    \includegraphics[width=0.276\textwidth, trim= 15mm 0mm 0mm 10mm]{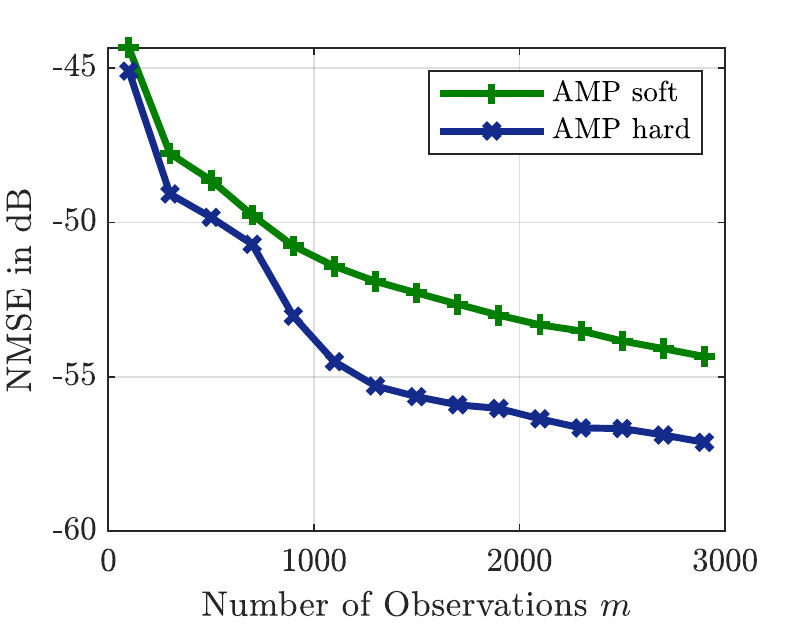}}
   	    \vspace{2mm}
	\subfigure[][]{
		\label{subfig:3}
   	    \includegraphics[width=0.276\textwidth, trim= 15mm 0mm 0mm 10mm]{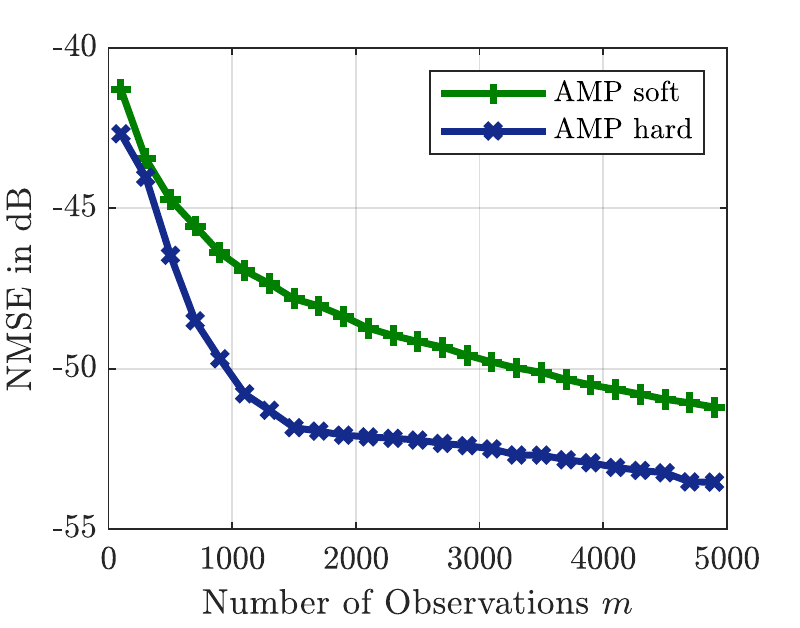}}
   	    \caption[]{
   	    \subref{subfig:softTh}~NMSE performance of the first THz subband out of $32$.
   	    \subref{subfig:hardTh}~NMSE performance of the third THz subband out of $16$.}
   	    \label{fig:nmsePerformance}

	\end{figure}
To circumvent the previously described problem for systems with low indeterminacy, a second
measurement series was conducted with a dynamically changing problem size $n$.
This is possible as systems with low indeterminacy $\delta$ are based on a smaller number of measurements $m$,
and therefore, the matrix $\mathbf{A} \in \mathbb{R}^{m\times n}$ allows larger $n$ while maintaining
an acceptable runtime of the algorithm. These measurements with a low indeterminacy are of high interest as
$l_0$-norm is a stricter measure of sparsity than $l_1$-norm, promising a better performance in
this setting. For the measurements, the indeterminacy was varied from $0.002$ to $0.215$ with a
step size of $0.003$ and the normalized sparsity was varied from $0.02$ to $0.4$ with a step size of $0.02$. The problem size was
set to $n = 50000$ for the smallest indeterminacy value of $\rho = 0.002$ and decreased to $n = 5000$ for
$\rho = 0.215$. The problem sizes were adjusted manually, such that the runtime was kept approximately
constant. The results are shown in Fig.~\ref{fig:translow}, where yellow patches indicate a high probability
of successful reconstruction and red patches show a low reconstruction probability. The figures show the performance of S-AMP and H-AMP and the theoretical phase transition for the $l_1$-norm
minimization is given as a blue line. As expected S-AMP matches the theoretical bound. On the other hand, H-AMP even outperforms the theoretical bound for $l_1$-norm minimization and
S-AMP for the low indeterminacy values used in these measurements. This can be explained
as follows. While for a high $\rho$ the obtained threshold $\tau$ lies within the area where the
algorithm does not converge, low values of $\rho$ lead to a high optimum threshold, that lies
within the area of convergence. Therefore the algorithm can converge with a better threshold
and outperform S-AMP.

\subsection{Observation for the THz Channel in the AMP framework}

As for THz channel estimation we have a very long THz channel impulse response and prefer to use short training sequences to sense the channel, the interesting region appears to be the one for low $\delta$ and low $\rho'$. Therefore H-AMP is of particular interest for THz channel estimation analysis.\par 
{
In the sequel, we assume BPSK symbols for the training sequence and therefore, the Toeplitz structured sensing matrix has unit $l_2$-norm normalized columns \cite{schramCS}. Furthermore, the THz channel was shown to be approximately sparse \cite{schramCS}. } 
As in \cite{schramCS} we consider a room of size $5\,\si{\m} \times 2.75\,\si{\m} \times 2.5\,\si{\m} $ for the indoor environment with a distance of $2.5\,\si{\m}$ between the transmitter and the receiver.
The corresponding \si{\THz} channel is divided into $N \in \lbrace 16,32 \rbrace$ subbands and the channel length is set to $n \in \lbrace 3223,1585 \rbrace$. The important parameters $k,\,m,\, \delta, \, \rho, \, \rho'$ are given in Table~\ref{table1}, with $N_u$ denoting the used subband and $[ x_{\text{begin}},x_{\text{end}} ]$ denoting an interval with start $x_{\text{begin}}$ and end $x_{\text{end}}$. All further relevant parameters are given in~{\cite[Table 1]{schramCS}}.
\par
Figs.~\ref{fig:nmsePerformance}~\subref{subfig:1} and~\subref{subfig:3} indicate the reached NMSE for the application of AMP to the THz channel. Clearly, an NMSE better than $-20$ dB can be achieved, identifying an appropriate success probability. Furthermore, it can be observed, that H-AMP outperforms S-AMP and it converges towards a very low NMSE. 
\par
From the parameters in Table~\ref{table1} arising for the THz channel and the results for the NMSE in Figs.~\ref{fig:nmsePerformance}~\subref{subfig:1} and~\subref{subfig:3} it can be inferred that for all considered $m$ a successful reconstruction will be achieved, as in Fig.~\ref{fig:translow} for low indeterminacy the yellow area can always be attained. Therefore, a successful channel estimation can be expected.
\begin{table}[b!] 
	\centering
	\caption{Characteristics of used subbands.}
	\label{table1}
	\begin{tabular}{|l|l|l|}\hline
		Subbands $N$              & $32$             & $16$             \\ \hline
		Considered subband $N_u$          & first & third \\ \hline
		Length of channel $n$        & $1585$           & $3223$           \\ \hline
	    Sparsity $k$      & $9$            & $9$            \\ \hline
		Length of training sequence $m$      & $[100,3000] $            & $[ 100,5000]$            \\ \hline
		Indeterminacy $\delta$      & $[0.063,1.893 ]$            & $[ 0.031,1.551]$            \\ \hline
		Compression rate $r$      & $[ 15.87,0.53]$            & $[ 32.26,0.64 ]$            \\ \hline
		Sparsity factor $\rho$      & $0.0057$            & $0.0028$            \\ \hline
		Normalized sparsity $\rho'$      & $[ 0.09,0.003 ]$            & $[ 0.09,0.002 ]$            \\ \hline
	\end{tabular}
\end{table}

\subsection{Simulation Results}\label{sub2}
The error metric used for evaluation of all AMP approaches is the $\mathrm{MSE}$ {in dB} defined as
\begin{equation}
\mathrm{MSE_{dB}}=\begingroup{10\log_{10}}\endgroup\Big\lbrack\dfrac{1}{R}\sum\nolimits_{r=1}^R \lVert \mathbf{h}-\hat{\mathbf{h}}^r \rVert_{2}^2\Big\rbrack,
\end{equation}
with $\hat{\mathbf{h}}^r$ being the estimated channel vector of the $r^{\mathrm{th}}$ of in total $R$ realizations. To calculate the MSE, the true channel vector is assumed to be known.
\par
\begin{figure}[t!]
    	\center{
   	    \includegraphics[width=0.45\textwidth, trim= 3mm 2mm 0mm 0mm]{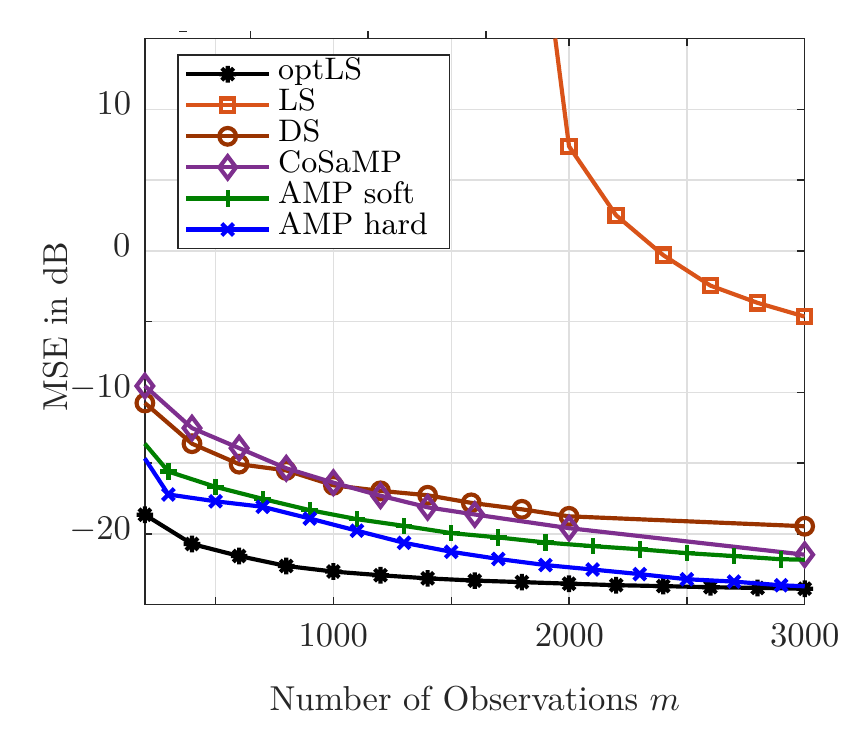}}
 	    \caption{Performance of different schemes for subchannel estimation for the first subband out of $32$ subbands.\label{fig:32}}
	\end{figure}
	\begin{figure}[t!]
    	\center{
   	    \includegraphics[width=0.45\textwidth, trim= 3mm 2mm 0mm 0mm]{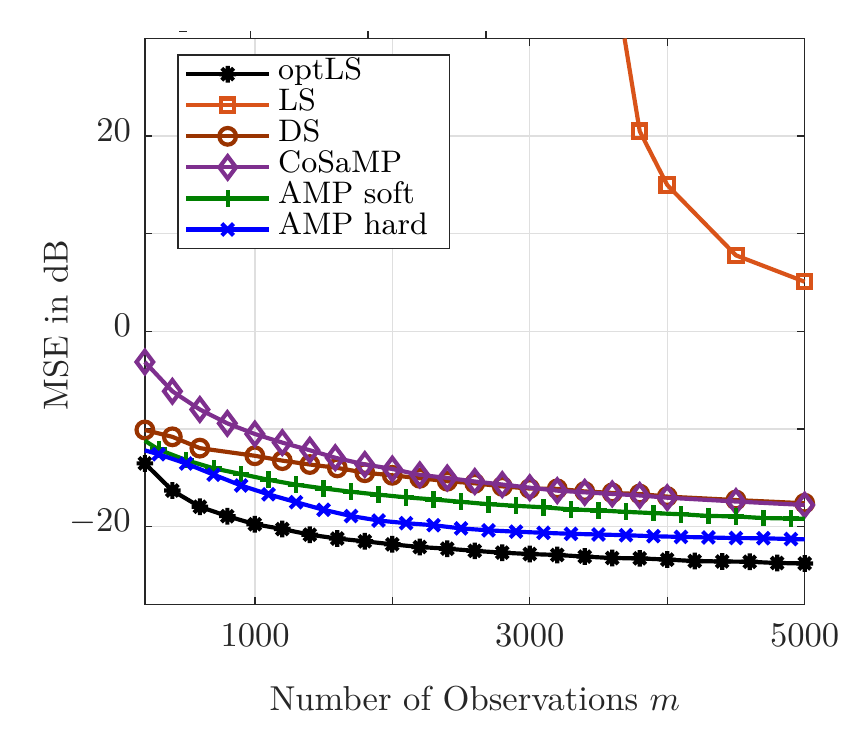}}
 	  \caption{Performance of different schemes for subchannel estimation for the third subband out of $16$ subbands.\label{fig:16}}
	\end{figure}
 We compare H-AMP with S-AMP and the methods studied in \cite{schramCS}. The results are illustrated in Fig.~\ref{fig:32} for the first subband out of $32$ subbands and in Fig.~\ref{fig:16} for the third subband out of in total $16$ subbands. Even for a low number of observations the AMP approaches outperform the Dantzig selector (DS), the compressive sampling matching pursuit (CoSaMP) method and conventional least squares (LS) estimation. H-AMP (AMP hard) outperforms S-AMP (AMP soft) and performs close to the oracle based benchmark (optLS), which is an upper performance bound for any realization scheme. It is observed that for cases with low indeterminacy in which the normalized sparsity is low, the H-AMP algorithm performs almost optimal.

\section{Conclusion}\label{V}
In this work, we have developed THz channel estimation algorithms based on H-AMP and S-AMP schemes.  Our investigations have demonstrated that for extremely sparse channel impulse responses, H-AMP outperforms S-AMP and tightly matches the optimal performance. Simulation results further revealed that the proposed AMP-based algorithms outperform all previously proposed CS methods and the conventional least squares approach for THz channel estimation.
\bibliographystyle{IEEEtran}
\bibliography{LiteratureAMP}
\end{document}